\documentclass[copyright,creativecommons,noderivs]{eptcs}
\providecommand{\event}{F-IDE 2019} % Name of the event you are submitting to
\usepackage{graphicx}
\graphicspath{ {./figs/} }

\usepackage[utf8]{inputenc} %useful to type directly diacritic characters
\usepackage{stmaryrd}

\usepackage[T1]{fontenc}
\usepackage{upquote}

\usepackage[svgnames]{xcolor}
\usepackage{fancyvrb}

\RecustomVerbatimEnvironment
  {Verbatim}{Verbatim}
  {fontfamily=courier,commandchars=\\\{\}}

\newcommand{\SYN}[1]{\Verb[formatcom=\color{DarkGreen}]"#1"}
\newcommand{\LEX}[1]{\Verb[fontseries=b]"#1"}
\newcommand{\VAR}[1]{\Verb[fontshape=sl]"#1"}
\newcommand{\VARSUB}[2]{\Verb[fontshape=sl]"#1#2"}
\newcommand{\FUN}[1]{\Verb[formatcom=\color{DarkRed}]"#1"}
\newcommand{\SEM}[1]{\Verb[fontshape=sl,formatcom=\color{blue}]"#1"}

\title{A Component-Based Formal Language Workbench}
\author{Peter D. Mosses
\institute{Delft University of Technology, Delft, The Netherlands}
\email{p.d.mosses@tudelft.nl}
}
\def\titlerunning{A Component-Based Formal Language Workbench}
\def\authorrunning{P.D. Mosses}
\begin{document}
\maketitle

\begin{abstract}
The CBS framework supports component-based specification of programming languages.
It aims to significantly reduce the effort of formal language specification,
and thereby encourage language developers to exploit formal semantics more widely.
CBS provides an extensive library of reusable language specification components,
facilitating co-evolution of languages and their specifications.

After introducing CBS and its formal definition,
this short paper reports work in progress on generating an IDE for CBS from the definition.
It also considers the possibility of supporting component-based language specification
in other formal language workbenches.
\end{abstract}

\section{Introduction}

Developers of major programming languages always give \textit{formal} specifications
of \textit{syntax}.
For \textit{semantics}, however, they usually resort to \textit{informal explanations}.
They sometimes define the formal semantics of sublanguages,
but scaling up to full languages is usually regarded as a huge effort, and not worthwhile.

% Moreover, programming languages evolve,
% mainly by extension with additional constructs.
% Co-evolution of syntax specifications with language extension is easy enough,
% generally requiring only local changes;
% but in many frameworks, semantic specifications may require major revision
% when a language is extended with significantly new constructs.
% Verbatim reuse of parts of previous language specifications is generally not possible;
% covert reuse by copy, paste, and edit is error-prone.

To encourage language developers to specify formal semantics of their full languages,
it is essential to reduce the effort required -- not only for an initial specification,
but also for co-evolution of language specifications with the specified languages.
The CBS framework aims to do just that,
by providing an extensible library of  \textit{reusable language specification components}.
The semantics of the components is defined once and for all,
so simply translating a programming language to compositions of components specifies the language semantics.
And specifying such translations can be significantly less effort than specifying language semantics directly.

Crucially, the definition of each component in the CBS library can be validated independently:
adding new components to the library cannot invalidate bisimulation equivalences of previous components.
After a component has been validated and released,
its defined behaviour cannot be changed,
so each occurrence of a particular component name in CBS specifications
refers to the same definition.
When a specified language evolves,
its translation to components always has to change accordingly,
as the components themselves cannot change.
% Partial operations can be extended to further arguments,
% and types can be extended to include further values,
% but existing values cannot be removed.

Use of CBS is supported by an IDE for editing and validating specifications of components and languages.
Validation is currently based on testing prototype implementations generated from specifications.
The IDE is itself generated from a formal definition of the syntax and static analysis of CBS,
which is specified in declarative meta-languages supported by the Spoofax language workbench \cite{Spoofax}.
CBS and its IDE have been developed by the PLanCompS project.%
\footnote{Programming Language Components and Specifications, \url{http://plancomps.org}.}

Here, after recalling the main features of the CBS framework (\S\ref{section:cbs}),
we give a progress report on the formal definition of CBS (\S\ref{section:cbs-spec}).
We explain how an IDE for CBS is generated from the definition (\S\ref{section:ide-gen}),
and how prototype implementations of programming languages are generated from
their specifications in CBS (\S\ref{section:interp-gen}).
We also compare CBS with some other language specification frameworks
regarding the possibility of defining libraries of reusable components
(\S\ref{section-related}).

\section{Component-based specification of programming languages}
\label{section:cbs}

We start by briefly recalling the main features of the CBS framework
for component-based specification.
For more detailed expositions, see
\cite{JLAMP,Modularity,TAOSD,JVLC}.

A CBS for a programming language is a specification of
an inductively-defined translation function,
mapping well-formed program phrases to terms formed from so-called
\textit{fundamental programming constructs} (`funcons').
The specification includes a \textit{grammar}
for the (concrete and abstract) syntax of the language,
and a \textit{translation equation} for each alternative of the grammar.
The translation of programs to funcon terms,
together with the semantics of funcons,
determines the semantics of the programs:
\begin{center}
        \includegraphics[width=0.8\textwidth]{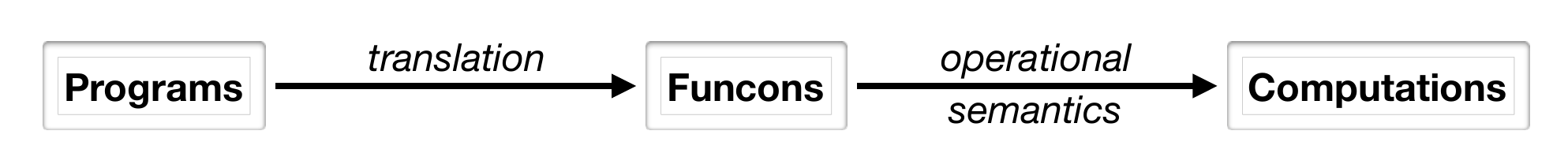}
\end{center}
The funcon definitions are reusable components of language specifications.

\paragraph{Funcons.}

A funcon represents a common and reusable computational concept,
such as variable assignment or looping.
It usually corresponds to a simple ingredient of constructs
commonly found in mainstream programming languages.
For example, an assignment in a program might be an expression,
returning either the target variable or the assigned value;
the funcon for assigning a value to a variable simply has that effect,
and does not return either of its arguments.
Loops in programs can often be terminated abruptly by break statements;
the funcons for loops simply propagate abrupt termination of their bodies,
and other funcons are used to handle abrupt termination.
% A declaration in a program might extend or override the current environment with further bindings;
% the funcon for a declaration simply returns the bindings,
% leaving it to other funcons to determine how they are to be combined with the current environment.
Funcons for integer operations return the unbounded mathematical results,
leaving it to other funcons to enforce boundedness.
Etc.

CBS uses a modular variant of structural operational semantics (MSOS) \cite{MSOS} to define funcons.
Elision of implicitly propagated semantic entities (environments, stores, etc.)\
ensures conciseness as well as modularity.
%, as in I-MSOS \cite{IMSOS}.
% CBS avoids the need for congruence rules by specifying argument strictness in funcon signatures.
Funcons are defined independently of any particular programming language,
and of each other.
Adding new funcons to the CBS library does not require any changes at all to previous definitions.
% This requires an exceptionally high degree of modularity.

% How funcons can be composed is determined by their signatures,
% independently of which funcons are used in their subterms.
% This makes the translations of different language constructs to funcons independent,
% which is important for supporting co-evolution.

The intended interpretation of funcon definitions in CBS is based on
their translation to MSOS,
and thereby as value-computation transition systems
\cite{FOSSACS2013}.
The translation from a precursor of CBS (MSDF) to MSOS was originally defined in Prolog
\cite{TAOSD};
funcon definitions are currently translated declaratively to monadic Haskell code
\cite{JLAMP},
where the monads involved correspond directly to the entities used in MSOS\@.
The definition of the translation is validated empirically by using generated code
to execute funcon terms
(manually-written unit tests, and translations of test programs in languages specified in CBS).

% CBS uses context-free grammars (with standard notation for regular expressions)
% to specify the abstract syntax of programming languages;
% the interpretation of such grammars as datatypes of ASTs is well established.
% To specify concrete syntax, CBS allows the use of SDF3 notation
% for associativity, (relative) priority, and follow-restrictions
%Note that the interpretation of disambiguation in SDF3 has recently been improved
% \cite[Ch.~2]{Eduardo}.

\paragraph{Syntax.}

CBS includes a variant of BNF context-free grammars,
extended with standard notation for regular expressions;
the interpretation of such grammars as datatypes of ASTs is well established.
Concise grammars for expression syntax are usually highly ambiguous;
CBS allows disambiguation (associativity, relative priority, etc.,
as in SDF3 \cite{SDF3}, \cite[Ch.~2]{Eduardo})
so the same grammar can be used to specify both abstract and concrete syntax.
% Alternative productions for the same nonterminal are grouped together,
% and AST constructors are left implicit.
Lexical syntax can be context-free (e.g., for nested comments).

\paragraph{Translation functions.}

The equations used in CBS to specify translations from program ASTs to funcons
look like semantic equations in denotational semantics:
the left side is an application of a translation function to an AST pattern,
with meta-variables matching sub-trees;
the right side is a funcon term containing applications of translation functions to meta-variables.
The meta-variables range over specified sorts of sub-trees.
Syntactic desugaring equations relate pairs of (composite) AST patterns.
The equations are interpreted inductively as functions on ASTs, as usual.

\begin{figure}[h]
    \centering
    % \fbox{A screenshot of the CBS IDE}
    % \vspace*{70mm}
    \includegraphics[width=0.93\textwidth]{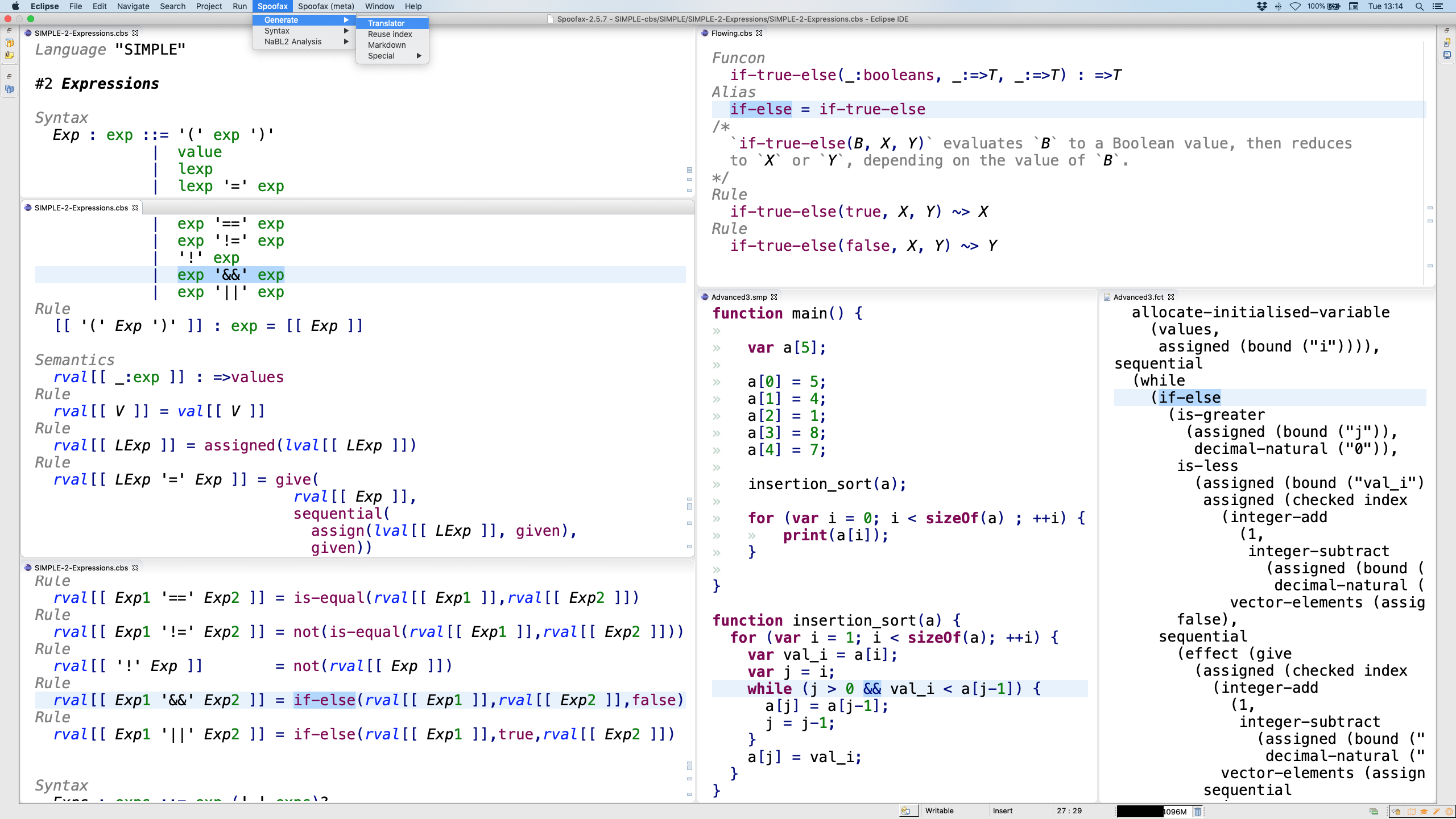}
    \caption{Using the CBS IDE on SIMPLE \cite[Languages-beta]{CBS-beta}}
    \label{fig:CBS-fragments}
\end{figure}

\bigskip\noindent
Figure~\ref{fig:CBS-fragments} illustrates the notation used in CBS for specifying funcons,
syntax, and translation functions.
% The signature of the funcon \FUN{if-true-else} (top right pane) implies that its first argument
% is to be pre-evaluated;
% the two rules specify its subsequent reduction to one of its two unevaluated arguments.
% Aliases such as \FUN{if-else} support formal abbreviation in funcon terms.
% The syntax for the sort \SYN{exp} (top left) introduces \VAR{Exp} as the stem of meta-variables
% ranging over phrases of sort \SYN{exp}.
% The rules for the translation function \SEM{rval} (bottom left)
% map ASTs of expressions of the form `\VARSUB{Exp}1 \LEX{\&\&} \VARSUB{Exp}2'
% to funcon terms formed from \FUN{if-else},
% the translations of \VARSUB{Exp}1 and \VARSUB{Exp}2, and the funcon \FUN{false}.
% The SIMPLE program (bottom centre) has been parsed and translated to a funcon term
% (bottom right) using the CBS IDE.
%
\begin{itemize}
    \item Top left: \VAR{Exp} is declared  as the stem of meta-variables
    ranging over \SYN{exp} phrases.
    \item Bottom left: The rules for the translation function \SEM{rval}
    map ASTs of \SYN{exp} phrases `\VARSUB{Exp}1 \LEX{\&\&} \VARSUB{Exp}2'
    to funcon terms formed from \FUN{if-else},
    the translations of \VARSUB{Exp}1 and \VARSUB{Exp}2, and \FUN{false}.
    \item Top right: The signature of the funcon \FUN{if-true-else} implies that its first argument
    is to be pre-evaluated;
    the two rules specify its subsequent reduction to one of its two unevaluated arguments.
    Aliases such as \FUN{if-else} support formal abbreviation in funcon terms.
    \item Bottom right: A SIMPLE program has been parsed and translated to a funcon term.
\end{itemize}

\section{Definition of the CBS meta-language}
\label{section:cbs-spec}

This section briefly explains how the syntax and static analysis of CBS
are defined \cite{CBS-Editor};
the meta-languages used are explained in the Spoofax documentation \cite{Spoofax}.

\paragraph{Syntax of CBS.}

We use SDF3 \cite{SDF3} to define the context-free syntax of CBS\@.
% In SDF3, the alternative forms for each sort of CBS construct are written as separate productions,
% as illustrated in Fig.~\ref{fig:cbs-syn}.
% The name given after a dot on the left of a production associates an AST constructor with the production.
% The form of lexical productions is illustrated in Fig.~\ref{fig:cbs-lex}.
% they construct leaves of ASTs.
% Layout and comments are implicitly allowed between symbols in context-free productions,
% but not in lexical productions.
% SDF3 normalises grammars to raw productions where the terminal symbols are single characters
% (or character ranges) and where the optional layout is made explicit.
SDF3 allows arbitrary context-free grammars;
disambiguation is supported by associativity and relative priority specification for context-free syntax,
and rejections and follow-restrictions for lexical syntax.

The current SDF3 definition of the syntax of CBS is given in
\cite[CBS/syntax]{CBS-Editor}.
Its formal interpretation is based on the transformation of SDF3 to SDF2
in \cite{Spoofax}
and the formal definition of SDF2 \cite{SDF2}.
The syntax definition has been empirically validated for coverage and disambiguation
against a collection of existing specifications written in CBS,
using a parser generated from it by Spoofax \cite{Spoofax}.
% It is also used directly in the toolchain of the IDE for CBS.

\clearpage

\paragraph{Static analysis of CBS.}

We currently use NaBL2 \cite{NaBL2} to define name resolution and type checking for CBS\@.
NaBL2 rules map ASTs to sets of constraints involving scope graphs.
The current NaBL2 definition of static analysis for CBS is given in
\cite[CBS/trans/static-semantics]{CBS-Editor}.

The rules for CBS name resolution require each funcon in the library to have a unique definition.
A~language specification can use any defined funcon
(without explicit import).
Grammar non-terminals, meta-variable stems, and names of translation functions are local to language specifications;
the variables used in a semantic equation or operational rule are local to it, and checked to be source dependent.

However, NaBL2 is not sufficiently expressive to define the intended type checking of funcon terms in CBS\@.
Our NaBL2 rules check that applications of funcons are consistent with signatures regarding number of arguments,
and whether arguments may compute undefined results,
but not the exact types of computed values
(which involves structural subtyping).
The rules also check whether the syntax patterns used in
specifications of translation functions match the specified grammar alternatives.
NaBL2 is to be superseded by a more powerful meta-language,
Statix \cite{Statix}, which should support full CBS type checking.

\section{Generation of the CBS IDE}
\label{section:ide-gen}

\begin{figure}[h]
    \centering
    \fbox{\includegraphics[width=0.98\textwidth]{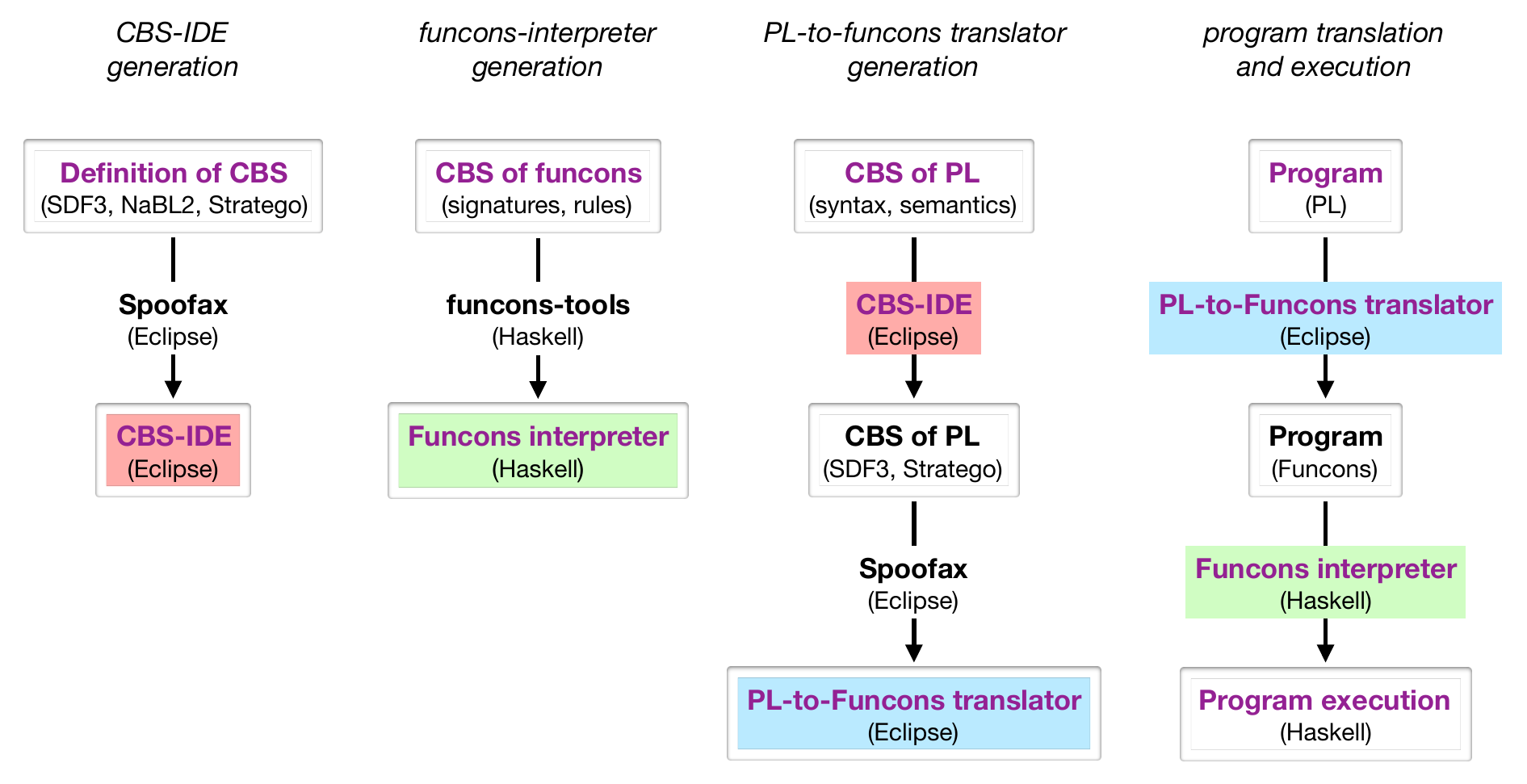}}
    \caption{Generation and use of the CBS IDE}
    \label{fig:cbs-tools}
\end{figure}

The IDE for CBS is implemented using Spoofax
\cite{Spoofax}.
As depicted in
Figure~\ref{fig:cbs-tools},
Spoofax generates the IDE directly from the definition of CBS:
it generates a parser for CBS from its SDF3 grammar
(with automatic error recovery, and syntax highlighting);
it generates name resolution and arity checking constraints for CBS
from the NaBL2 rules.
Parsing and static analysis errors in  specifications are flagged by Eclipse.

Each language specification in CBS is a separate Eclipse project,
with shared access to the funcons library,
which is stored in a remote repository.
A language specification can be split into
% (optionally numbered and hierarchical)
sections,
and stored in any number of files.
Multi-file name resolution generates hyperlinks from names to their definitions.
% funcon names are global,
% whereas syntactic sorts, translation functions, and meta-variables
% are local to the enclosing language specifications.
% Paths to files are never mentioned in CBS specifications,
% so the distribution of items between files is irrelevant.

\section{Generation of language interpreters from CBS}
\label{section:interp-gen}

The generated IDE for CBS has a menu to generate various artefacts
when editing a language specification written in CBS:
a syntax-aware editor to parse programs and translate them to funcons,
a list of reused funcons,
and Markdown pages with embedded HTML for hyperlinks and highlighting.
When a language specification is changed,
stale generated files are regenerated.
When the generated Markdown pages are uploaded to GitHub Pages,
they create a responsive website (e.g., \cite{CBS-beta})
where language specifications and funcon definitions can be browsed
using hyperlinks for navigation, as in Spoofax.

The Haskell package \textit{funcons-tools} \cite{Funcon.Tools} can be used as an external tool
from the CBS IDE\@.
It generates highly modular Haskell code for funcon interpreters from CBS definitions of funcons \cite{JLAMP}.

% The generated editor files are in SDF3 and Stratego.
%
% The list of funcons that are reused in a language specification is obtained by
% filtering the full index of funcons using name resolution data.

\section{Related work}
\label{section-related}

CBS appears to be the only framework that currently provides a library of reusable components
of programming language specifications.
If the CBS library could be specified in other frameworks,
it might encourage users of those frameworks to exploit funcons,
and provide further tool support for CBS users.

The K-framework \cite{K} is highly modular,
and has been used to specify the semantics of several major languages.
The funcons used in the CBS specification of SIMPLE
\cite[Languages-beta]{CBS-beta}
have already been re-specified in K,
allowing the CBS of SIMPLE programs to be run with the K tools
\cite{FunKons}.
However, the specification of the K configurations was monolithic,
and depended on the set of funcons.

Reduction semantics is a popular form of operational semantics,
with mature tool support including Redex
\cite{Redex}.
Language extensions can be specified smoothly in Redex,
but it is unclear how to define an open-ended collection of funcons:
reduction semantics requires grammars for evaluation contexts,
and the evaluation contexts for a particular funcon seem likely to depend
on which other funcons are needed.

XASM \cite{XASM} is a component-based language for Abstract State Machines \cite{ASM},
supporting the use of Montages \cite{Montages} for specifying programming languages.
However, it appears that the original home page
(\href{http://xasm.org}{\texttt{xasm.org}})
and the subsequent SourceForge project
(\href{http://xasm.sourceforge.net}{\texttt{xasm.sourceforge.net}})
are no longer in use,
and that no library of reusable XASM components has ever been released.

The Overture F-IDE \cite{Overture} supports dialects of
the Vienna Development Method (VDM)\@.
Those specification languages contain many constructs which, like funcons,
correspond closely to constructs of high-level programming languages:
assignment statements, while-loops, exception-handling, etc.
Thus translating a programming language to a VDM dialect would be
similar to specifying its CBS\@.
It could be interesting to investigate defining
an extensible library of funcons in VDM, with IDE support in Overture.

\section{Conclusion}
\label{section-conc}

The syntax and static semantics of CBS have been formally defined in \cite{CBS-Editor} using
the declarative meta-languages SDF3 and NaBL2.
% Disambiguation specification in SDF3 has recently been improved,
% and is to be embedded in CBS.
The NaBL2 definition of CBS type checking
needs to be reformulated in the new Statix meta-language \cite{Statix},
to specify rigorous subtype checks,
after which the IDE generated from the definition of CBS is to be released
as an Eclipse plugin.
% The responsiveness of the IDE should be significantly improved
% by the incremental implementation of Statix analysis.%
% \footnote{The version available at \cite{CBS-Editor} is merely a proof of concept.}

Current examples of language specifications in CBS
include OCaml-Light;
scaling up to full OCaml (and other major languages)
remains to be demonstrated.
Bisimulation properties of funcons can be proved once and for all \cite{FOSSACS2013}; the proofs could be included with the  definitions in the library.

The CBS development is part of the PLanCompS project.
The project was initially supported by an EPSRC grant (2011--16),
and it is continuing as an open collaboration;
new participants are welcome.

\bibliographystyle{eptcsini}
\bibliography{f-ide-short}
\end{document}